# High Finesse Buckled Microcavities

S.W. Ding[1*†], B. Grinkemeyer[2*], G.E. Mandopoulou[2], R. Jiang[1], A.S. Zibrov[2], G. Huang[1], K. Yang[1], M.D. Lukin[2], M. Lončar[1]

[1] *John A. Paulson School of Engineering and Applied Sciences, Harvard University, Cambridge, Massachusetts 02138, USA*
[2] *Department of Physics, Harvard University, Cambridge, Massachusetts 02138, USA*
[*]*These authors contributed equally to this work.*
[†]*Present address: AWS Center for Quantum Computing, San Francisco, CA, USA (work done prior to joining Amazon)*

## Abstract

Optical cavities are widely used in modern science and technology to enable a wide range of both quantum and classical applications. Recently, the growing demand for miniaturization and high performance has fueled the exploration of new fabrication methods beyond traditional polishing techniques and macroscopic mirrors. Visible and near-infrared (NIR) wavelengths are particularly important for quantum applications, where achieving low-loss resonators is also more challenging than in the telecom range, presenting unique challenges and opportunities for microscopic cavity systems. Here, we present a novel fabrication method for making NIR microcavities using buckled dielectric membrane mirrors, achieving a record finesse of 0.9 million at 780 nm. We demonstrated flexible device geometries – including singular mirrors and mirror arrays, featuring radii of curvature ranging from 1 mm to 10 mm. The fabrication process offers high uniformity, high yield, and robust performance across a wide range of cavity lengths. Additionally, we can produce easy-to-assemble microcavity packages, with a total volume of ~2 (4) mm³, featuring optical modes with a linewidth of 5.16 MHz (570 kHz) and a free spectral range (FSR) of 3.18 THz (150 GHz). Our results extend the frontier of microcavity fabrication for classical and quantum photonic technologies.

## Introduction

The Fabry–Perot (FP) cavity is one of the most versatile and widely utilized optical devices. The high-quality factors and frequency stability that state-of-the-art FP resonators offer enable significant advancements across various fields, including cavity quantum electrodynamics (cQED)[1,2], quantum information science[3–5], high-precision sensing[6–8], and timekeeping[9,10]. Traditionally, the high-performance mirrors forming these cavities are manufactured using super-polishing techniques and low-loss dielectric coatings[11]. However, their large size and radii of curvature (ROCs) make them unsuitable for compact and scalable optical systems. Additionally, these ROCs also result in larger mode volumes, which limit their performance for certain quantum applications[3–5]. In recent years, motivated by the need for microscopic devices that enable photonic integration and efficient quantum light-matter interfaces, several microfabrication techniques, including laser machining and chemical etching, have been developed to address these limitations and produce high-quality micro-mirrors[12–14].

Figures of merit, such as finesse, waist size, form factor, quality factor ($Q$), and mode matching/coupling efficiency, can be used to characterize the performance of an FP cavity. Depending on the applications, certain parameters become more important than others. For example, for a highly stable frequency reference[10], the absolute linewidth and its long-term stability are critical. In this case, the cavity design needs to be optimized to balance stability while maintaining a narrow linewidth, which usually leads to a large cavity mode volume along with an extremely narrow cavity linewidth. In contrast, in systems designed

to enhance light-matter interactions[5], the finesse and the waist size are more important, while a moderate cavity linewidth is preferred for larger bandwidth, which tends to favor smaller cavities. The mode matching/coupling efficiency also plays a role in systems with limited photon rates. In general, among these figures of merit, finesse ($F$) is one of the most fundamental, as it is a direct reflection of the loss in the resonator. It is defined as $F = 2\pi/\mathcal{L}$, where $\mathcal{L}$ is the round-trip loss in the resonator. In many resonators, the primary limiting parameter is the optical scattering loss ($S$), dominated by the surface roughness ($\sigma_s$) of the mirror surface, particularly for FP cavities, due to the fabrication process. Quantitatively, the finesse limit of a resonator is inversely related to the surface roughness by $F \propto 1/S \propto 1/\sigma_s^2$ [15].

Fig. 1(a) plots the quality factor ($Q$) against the normalized cavity length ($L/\lambda$), for a variety of representative cavities related to this work. Here, $L$ is the cavity length and $\lambda$ is the wavelength. The dashed lines represent finesse-constant values: $F \sim Q/(L/\lambda)$. The microcavity data points compare favorably with state-of-the-art works, including isotropically etched silicon micromirrors (Si)[13], resist-reflowed glass/ULE mirrors (Reflow)[9,14], and CO2 laser-ablated fiber cavities (Fiber)[12]. For broader comparison, we also include representative data points from other major types of optical cavities, such as traditionally polished macroscopic mirrors (Traditional)[11], integrated photonic ring resonators (Integrated)[16–18], and toroidal microdisk resonators (Microdisk)[19–21], to reflect their best demonstrated performance. These cavity types each offer unique trade-offs and play important roles across a broad range of photonic applications.

On one hand, the traditional approach based on polished mirrors has the highest $F$ and $Q$ and remains the preferred option for applications that require narrow linewidth and high stability[10,11]. However, such cavities are large, and scalable fabrication is challenging. On the other hand, microfabricated FP cavities have been produced by various scalable methods. Our work (indicated with stars) features state-of-the-art $F$ achieved in small cavities fabricated using a simple fabrication process and the buckling of the dielectric coating. Other microfabrication methods, such as isotropic etching of Si and CO2 laser ablation, can produce hundreds of cavities with smaller sizes and slightly lower $F$ and $Q$ values than the approach presented here; they are frequently used in quantum applications as photonic interfaces due to their small ROCs resulting in small mode volumes and strong light-matter interactions[5,12,13,22]; reflow-based FP micro cavities can span a large range of ROCs and achieve slightly higher $F$ and $Q$ to those demonstrated in this work, but require more careful etch calibration[14]. These microcavities have found applications in compact laser applications due to their very narrow linewidths[9], but their larger ROC makes them less suitable for quantum applications. Integrated photonics microcavities based on ring resonators (typically in SiN and SiO2 material platforms) are compatible with large-scale manufacturing and integration, but tend to have lower $Q$ and $F$, limited by material absorption and roughness-induced scattering. They are currently used to realize compact lasers, modulators, and many other integrated photonic applications and innovations[16–18,23]. Microdisks and toroids feature some of the highest $F$ and $Q$ values (e.g., when made in $SiO_2$, $CaF_2$, or $MgF_2$ platforms), and can be small and microfabricated; however, they usually require prisms, tapered fibers, or other optical components for coupling, which makes them unsuitable for integration. Still, they are widely employed in commercial narrow-linewidth lasers and used for advanced optical applications[19–21].

The rise of quantum optical applications has driven the development of optical cavities in the near-infrared and visible ranges, where achieving low surface roughness is more critical than at telecom wavelengths. Specifically, it calls for a fabrication method tailored to the new wavelength range while maintaining state-of-the-art performance. Importantly, shorter wavelength imposes stricter requirements on the surface quality (roughness, e.g.) to achieve ultrahigh finesse, since $S$ is inversely proportional to $\lambda^2$—specifically, $S = \left(\frac{4\pi\sigma_s}{\lambda}\right)^2$ [24]. Therefore, exploring shorter-wavelength microcavities also pushes the boundary of nano- and microfabrication techniques to produce smooth surfaces.

In this work, we present a simple, robust, and scalable method for fabricating state-of-the-art visible microcavities with ultra-smooth surfaces using buckled dielectric membrane mirrors, as shown in Fig. 1(b). These micro-cavities combine high finesse, standard Si fabrication, small mode volume, and small ROC, making them ideal candidates for applications in visible lasers[25,26], sensing[6–8,27], and quantum networking[3,4,28].

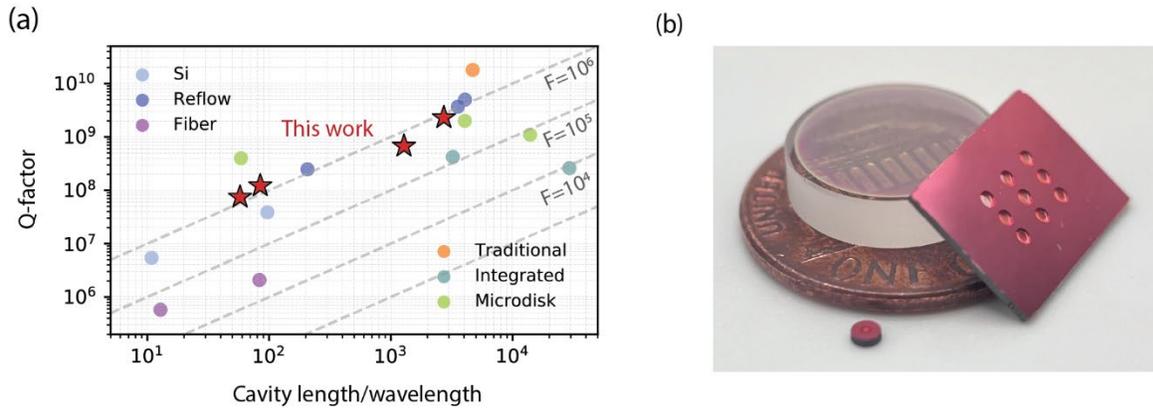

Fig. 1. Compact micromirror array and device (a) Finesse and quality factors as figures of merit for micro optical cavities: isotropically etched silicon micromirrors (Si)[13], resist reflowed (Reflow)[9,14], CO2 laser ablated fiber cavities (Fiber)[12], traditionally polished macroscopic mirrors (Traditional)[11], integrated photonic ring resonators (Integrated)[16–18], toroidal microdisk resonators (Microdisk)[19–21], and this work (red stars). (b) a photo of the fabricated micromirror array and a micromirror next to a penny and a half-inch diameter Thorlabs mirror used in a commercial Fabry Perot cavity.

## Result

### Mirror fabrication and characterization

Our fabrication approach is based on standard nano- and microfabrication tools and methods, as shown in Fig. 2(a). We start with a commercially available 4" silicon wafer that is double-sided polished. The RMS surface roughness ($\sigma_s$) of an as-acquired wafer is around 0.3 nm (Fig. 2(a)(1)). To improve the smoothness, we grow 2.5 µm of wet oxide and remove it with 49% HF (Fig. 2(a)(2) and (3)). This step is repeated two to three times, resulting in a surface roughness of ~50 pm down to 33 pm (measured on a randomly selected 2 µm x 2 µm area on the wafer), using atomic force microscopy (AFM). Next, a dielectric coating consisting of 21 pairs of quarter-wave stacks of $SiO_2$ and $Ta_2O_5$ layers, in addition to a final protective capping layer of a half-wave stack of $SiO_2$ is deposited on the wafer using ion beam sputtering (Fig. 2(a)(4)). This step is performed by FiveNine Optics. The total thickness is around 5 µm. The coating is quoted to have 1.37 ppm loss at 780 nm and gives the wafer its magenta red color (see SI). Next, the coating is protected by PMMA, and photolithography using SPR photoresist (thickness of 7 µm) is used to define the mirror shape on the back side (Fig. 2(a)(5)). Finally, the dielectric stack is suspended by etching through the Si chip/wafer from the backside using a Bosch etch process with $SF_6$ and $CF_4$ (Fig. 2(a)(6)), resulting in a freestanding mirror. The sample is cleaned using PG remover.

The suspended mirrors are naturally buckled due to the built-in compressive strain in the dielectric coating. As shown in Fig. 2(b), an array of micromirrors with a diameter of $d$ = 600 μm can be created with high yield and uniformity.

The shape of the resulting mirror can be simulated with high accuracy, as shown in Fig. 2(c), with an $R^2$ value of 99.98%. The simulation is conducted using COMSOL, assuming a compressive strain of 0.3% in the film and a fixed boundary condition on the circular edge (see SI). The shape cannot be described analytically using a simple functional form, such as a Gaussian or Bessel function. To explore the range of geometries that can be fabricated using this approach, we vary the diameters of the suspended mirrors and can achieve reliable fabrication for ROCs ranging from .9 mm to 21 mm.

Finally, to ensure the high finesse of the cavities using two microfabricated mirrors, the mirror shape needs to be rotationally symmetric. We use an optical profiler to infer the shape of our mirrors and find that the two orthogonal cross-section profiles are identical ($R^2$ = 100.00%), as shown in Fig. 2(e).

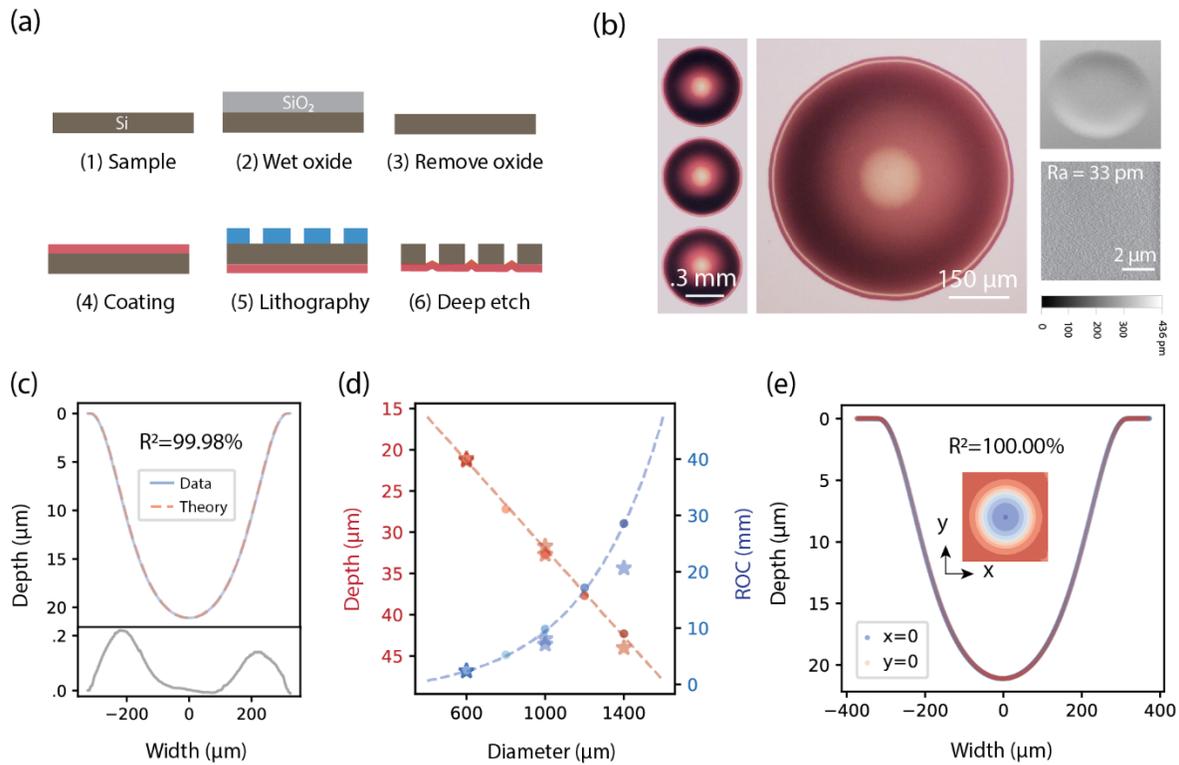

Fig. 2. Micromirror fabrication and profile characterization (a) Fabrication flow diagram. (1): The fabrication starts with a Si wafer or chip. (2)(3): then, the sample is subjected to "wet thermal smoothing", which turns outstanding surface features into oxide and removes them. (4): dielectric layers are deposited onto multiple wafers at once. (5)(6): openings are defined using lithography and etched through the Si substrate to allow the strained film to buckle. (b) The fabricated micromirrors have a diameter $d$ = 600

µm. The two panels on the left are optical microscope images, while the top right image is an SEM image. The bottom right image is an AFM image of the Si substrate before the dielectric stack deposition, representing the lower bound of the mirror surface roughness. (c) The simulated mirror shape, assuming a uniform strain of .3% in the film, compared to the measurement, shows almost perfect agreement. The inset is a contour plot of the mirror profile. (d) The depth and ROC for different fabricated mirrors as a function of the lithographically defined mirror diameter. The circular points indicate simulated ROCs and depths, and the stars correspond to measured values. Notice the stars are multiple data points overlapping perfectly for each defined diameter. (e) The fabricated mirror's X and Y cross-section profiles show perfect rotational symmetry. The profile data shown in (c) and (e) are taken by an optical profiler on a $d = 600$ µm mirror.

## Finesse measurements

To form a microcavity, two mirrors are mounted and aligned on translation stages. One mirror is attached to a piezoelectric transducer, which controls the cavity length across a full free spectral range (FSR), as shown in Fig. 3(b). The micromirrors are characterized using a swept-ring-down method[29–31], in which the cavity length, and therefore resonance, is rapidly modulated while a narrowband laser, locked to a rubidium resonance at 780.24 nm, is sent through the cavity. As the cavity sweeps through the laser wavelength, the photodetector collects the transmission data. The detector is connected to an oscilloscope to visualize the ring-down measurement. To ensure that we measure the TEM00 mode of interest, we also monitor the cavity output with a camera. This allows us to confirm the mode purity and thus optimize the alignment. This configuration facilitates measuring high finesse and quality factors at different cavity lengths (see SI).

The FSR measurement is shown in Fig. 3(b). The peaks correspond to different modes: the patterns show the light intensity distribution of the mode. To extract the $F$ and $Q$ of the microcavity, we zoom in on the TEM00 peaks and fit the energy decay to an exponential profile. The decay curves that feature the highest $F$ and the highest $Q$ are shown in Fig. 3(d) and (e), respectively. For the highest finesse point, we find $\kappa = (2\pi)\, 233 \pm 1$ kHz which translates to $F = 0.89 \times 10^6$. For the highest Q point: $\kappa = (2\pi)\, 124 \pm 0.5$ kHz, which translates to $Q = 2.3 \times 10^9$.

The fringes in the ring-down signal are a result of light built up in the cavity leaking out as the length changes and interferes with itself, and are signatures of a high $Q$. The functional form is a convolution of an exponential decay and a complementary error function, with a constant offset:

$$I \propto e^{-\kappa t} |erfc(\frac{(i+1)\,\kappa/2 - i\alpha(t-t_0)}{i\sqrt{\alpha}})|^2$$

Here, $\kappa$ is the full-width half max (FWHM) of the cavity, and $\alpha$ is a parameter that reflects the velocity of the cavity frequency sweep. Both parameters are fitted to extract $F$ and $Q$. In our measurements, $\alpha$ varies with the length of the cavity ($L$) due to the change in frequency being proportional to the FSR; for our measurements, typical values range from 1.5 THz/ms to 75 GHz/ms. Consequently, $\alpha$ can be used to extract the length of the cavity together with the FSR in time ($\Delta t_{FSR}$) (see SI). $F$ is obtained by:

$$F = \frac{\lambda Q}{L} = \frac{2\alpha \Delta t_{FSR} \lambda Q}{c}$$

An alternative method to extract $L$ uses the mode spacing (between the TEM00 and TEM01 e.g.) and measured ROC of the mirrors (see SI). We corroborate both measurements to determine the cavity length precisely, accounting for nonlinearity in the piezo and other errors, especially for short cavity lengths that cannot be measured accurately physically.

We study $F$ and $Q$ corresponding of the microcavity at different lengths $L$, from 66 to 3750 μm(Fig. 3(c)). The finesse monotonically drops as the cavity becomes longer due to diffractive loss, related to the shape deviating from a sphere, and other imperfections seen by the light. The $Q$ increases at the beginning and then plateaus, as both $F$ and the $L$ change linearly, before dropping as the cavity length significantly exceeds the ROC. This trend is captured in Fig. 3(c) and can be modeled qualitatively (see SI).

The finesse of a Fabry Perot cavity is dictated by loss in the system as $F = \pi/(T + S + A)$, where $T$, $S$, and $A$ represent the transmission, scattering, and absorption loss per mirror, respectively. $T$ is dictated by the dielectric stack, which can be designed to be more or less reflective depending on the application; $S$ originated from surface roughness or unideal mirror shapes, which fundamentally sets the limit of a cavity's finesse, usually as a result of a certain fabrication process; $A$ is primarily due to the materials used for coating. Our mirrors have quoted $T \geq 1.33$ ppm at 781 nm, which is obtained by a transmission measurement (see SI). The measured cavity finesse $F_{780\ nm} = 0.89\times 10^6$ can bound the rest of the loss: $A + S \leq 2.20$ ppm. If we assume that there is no absorption loss, $A = 0$, we can find an upper bound on the surface roughness, $S = \left(\frac{4\pi\sigma_S}{\lambda}\right)^2 \leq 2.20$ ppm, which translates to a surface roughness of $\sigma_S \leq 90$ pm. This value is slightly higher than our measured surface roughness because the coating has an intrinsic nonzero $A$. It is challenging to distinguish scattering loss from absorption loss. According to the vendor, a typical $A$ value for this specific coating is between 1.5 ppm and 2 ppm, which means $S$ achieved by our method is comparable, if not less than, state-of-the-art visible dielectric coating loss[11]. Notice, $S \propto 1/\lambda^2$, which is one of the fundamental reasons why it is more challenging to achieve high finesse FP cavities at a lower wavelength with the same fabrication process and the resulting surface roughness. For example, since the 780 nm wavelength of interest here is roughly half of 1550 nm, typically considered in cavity applications, the surface needs to be four times smoother (assuming that state-of-the-art mirrors are roughness-limited) to achieve the same finesse or scattering loss at the two wavelengths. In other words, using the measured roughness ($S \sim 2.20$ ppm) and loss values ($T \sim 1.33$ ppm) in our current system, we can extrapolate the performance of these cavities at telecom for comparison with other systems, which yields a finesse of $3.5 \times 10^6$. The microcavities shown in this work and the corresponding fabrication process therefore feature state-of-the-art surface roughness, which may be explored for applications beyond FP cavities.

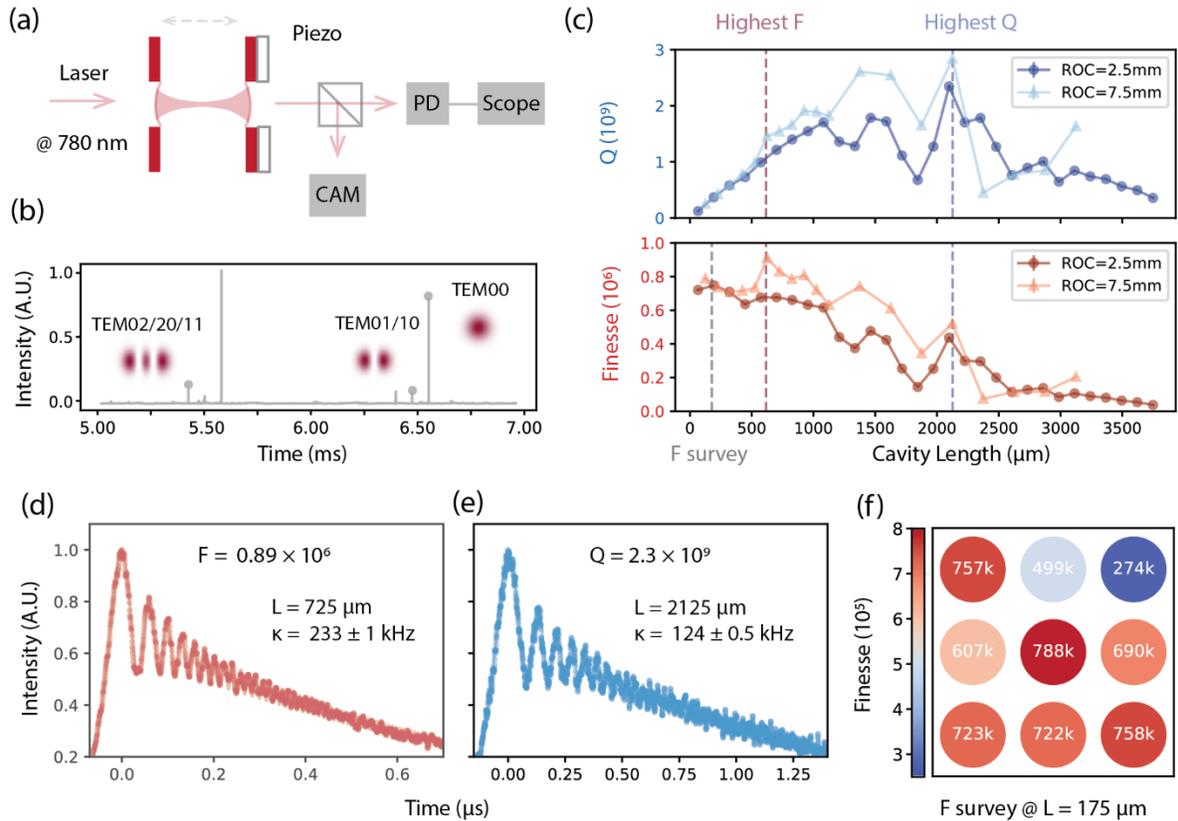

Fig. 3. Microcavity characterization (a) A diagram of the measurement setup. CAM: camera; PD: photodetector; Scope: oscilloscope. In our approach, the transmission of the cavity is measured as the cavity length is scanned using a piezoelectric transducer. (b) The FSR of the cavity at a cavity length of 58.7 µm is 2.5 THz. There are three modes visible in the range. The modes are identified by the shape of the transmitted light, as imaged using the camera, and by comparing them to the Hermite polynomials. (c) The Q and F vs. cavity length data show a finesse fall-off as a function of distance, which we attribute to diffractive losses due to the shape of the mirror. The finesse does not drop linearly with length; as a result, we find that Q continues to increase to a maximum, where F begins to fall more quickly. (d) The data point that corresponds to the highest finesse of 0.9 million. (e) The data point that corresponds to the narrowest linewidth of $(2\pi)$ 124 ± 0.5 kHz. (f) The F survey to show the yield on a 3-by-3 chip of $d$ = 600 µm mirrors. They are measured against another $d$ = 600 µm buckled micromirror. Most of them show high finesse at $L$ = 175 µm, which is consistent with the data points from (c), as indicated by the reference grey dashed line.

## Integrability

The fabrication method also enables the creation of small, assembled cavities with a size that can be integrated into a fiber[27] or an integrated photonic chip[9]. We demonstrate this by creating a compact cavity system using the mirror cutouts of the silicon chip, with only heat-cure glue and manual alignment, which can be easily implemented in a lab setting.

Individual mirror chips are separated from the wafer by etching through the Si substrate during the same etch step used to create the suspended mirrors. In this work, we choose the outer diameter of the chip to be

1.7 mm (Fig. 4(a)). The two chips are glued together to produce a cavity with a narrow resonance, with or without a spacer used between the two mirrors. We created two cavity assemblies with different lengths (Fig. 4(a) inset). The shorter cavity stack is 1 mm long, and the cavity is formed by the two concave mirror surfaces pressed against each other, forming a short cavity of 45 μm. The longer cavity has a 1 mm spacer, resulting in a total cavity stack length of 2 mm. The cavity is measured by directly placing a cleaved fiber against the back of the mirror and collecting the transmission, as shown in Fig. 4(a), for the short cavity. The measured linewidth is $\kappa = (2\pi)\ 5.16 \pm 0.14$ MHz, with a very large FSR of 3.18 THz, which makes it ideal for compact laser locking applications [19,32–34]. The figures of merit of this microcavity assembly are measured to be: $F_{45\ \mu m} = 0.616 \times 10^6$, $Q_{45\ \mu m} = 7.45 \times 10^7$, $F_{1\ mm} = 0.263 \times 10^6$, $F_{1\ mm} = 6.74 \times 10^8$, which are consistent with the previous finesse versus length measurements. The minor drop in finesse is due to the misalignment of a manual gluing process, as shown in Fig. 4(a), which also highlights the robustness of this assembly.

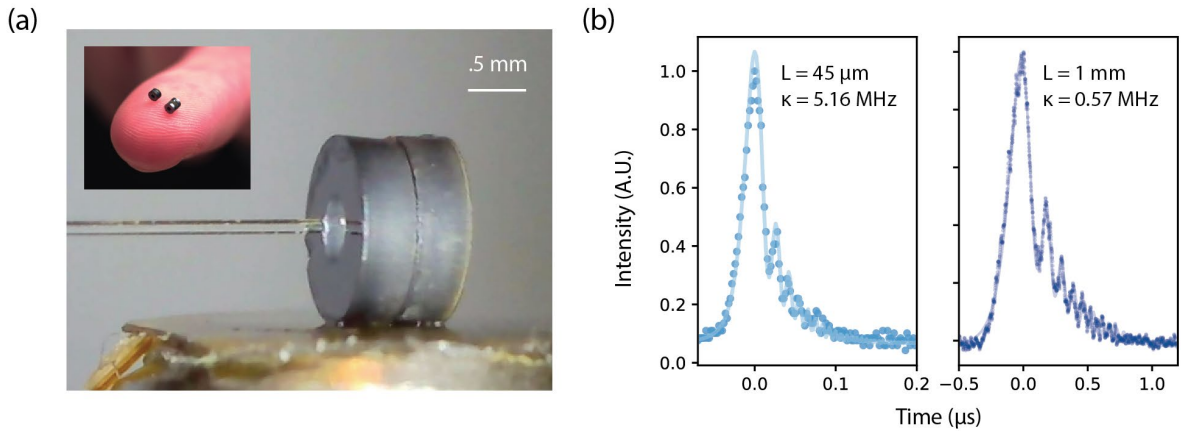

Fig. 4. Microcavity assembly (a) A photo of the measurement setup and the actual measured cavity device. The fiber used is 780HP, suitable for operation in the 780-970 nm wavelength range. The inset shows the cavities with two different lengths (45 μm and 1 mm), placed on the finger tip. The cleaved fiber tip is inserted into the cavity mirror slot from the back. (b) The measured ringdown of the cavity for the cavity assembly at different lengths. In this case, the laser frequency is swept.

## Summary and Conclusion

We demonstrate a novel method of producing state-of-the-art micromirrors and microcavity systems. The fabrication is simple, robust, flexible, and scalable, and can be achieved in a common cleanroom environment with high tolerance to errors. The resulting devices can be single mirrors, arrays of mirrors, or cavity assemblies, all characterized by robust performance and flexible configurations.

The microcavity described in this work can play an important role in both quantum and classical optics. The combination of small device size (down to 2 mm³) and moderate ROC enables integration with existing architectures for neutral atoms and ion trap systems as an optical interface for quantum networking. Specifically, this cavity geometry offers cavity lengths and small diameters that would provide ample optical access for neutral atom and trapped ion experiments. Furthermore, the long cavity lengths would mitigate the detrimental effects of the dielectric surfaces on the qubits[35,36], while still maintaining a small mode volume for strong light-matter interactions[37,38]. Achieving this regime is challenging using mirrors produced by conventional polishing or ablation methods, which typically yield ROCs that are either too large or too small. In classical applications, these microcavities can be used as high-efficiency filters, transfer cavities, or external resonators. For example, due to their small size, these optical cavities can be tuned with high-bandwidth piezoelectric modulators, allowing for stabilization against acoustic noise, which is challenging for larger mirrors[39], and providing robust stability to their state-of-the-art linewidth. The stabilized cavity can then be utilized as a compact, narrow filter, an external cavity for a laser, or a transfer cavity for stabilizing lasers of vastly different wavelengths.

## Funding information

This work was supported by the US Department of Energy (DOE) QSA Center (DE-AC02-05CH11231), the National Science Foundation (NSF) (grant PHY-2012023), the Center for Ultracold Atoms (an NSF Physics Frontiers Center), and AFOSR MURI FA9550-23-1-0333 (via U Colorado). G.H. acknowledges financial support from the Swiss National Science Foundation (Postdoc Mobility, grant number 222257).

## Acknowledgement

This fabrication work was performed at the Center for Nanoscale Systems (CNS), Harvard University. The coating is performed by FiveNine Optics. We thank Tamara P. Zibrova for discussions and help with mirror coating.

# Supplementary Information

# High Finesse Buckled Microcavities

**Properties of the dielectric coating**

The dielectric mirror coating is performed by FiveNine Optics, using ion beam sputtering. Based on measurements provided by the vendor, the center wavelength is $\lambda = 781$ nm. Each mirror consists of 21 quarter-wave thick layers formed by alternating materials of different refractive indices (n): $SiO_2$ ($n \sim 1.47$) and $Ta_2O_5$ ($n \sim 2.10$). The total thickness of the coating is therefore 4.74 µm, and the transmission (loss) is measured to be $T = 1.33$ ppm (Fig. S1). This sets the lower limit for cavity losses to 2.66 ppm per round trip. This value does not include absorption and scattered losses in the coating. It is difficult to distinguish absorption loss from scattering, thus our measurements can only bound the scattering loss by assuming no absorption loss. FiveNine Optics estimates absorption losses for our mirror as 3-4 ppm at 780nm by extrapolating from absorption losses at nearby wavelengths. However, based on our measurements, it is clear that our coating significantly outperforms this estimate as we find our losses to be below 3ppm. This also suggests that our scattering loss can be significantly lower than the bound we measure.

The shape of the mirror can be simulated almost perfectly by COMSOL, given the coating thickness and materials, where we treat the compressive strain as uniform and a parameter to adjust for. The estimated strain is .3%.

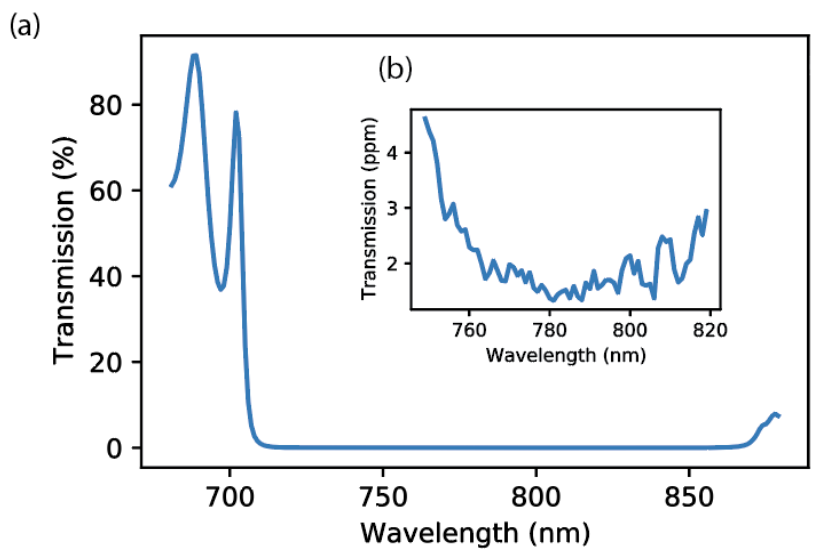

Fig. S1. Transmission (loss) of the mirror provided by the vendor. The measurement is performed on a reference glass sample produced in the same coating run as the wafer we used in this work. This measurement contains both absorption and scattering loss inherent to the coating. (a) Transmission profile of the coating. (b) A zoomed-in plot to show the high reflectivity region of the transmission window, where the lowest transmission loss is 1.33 ppm at around 781 nm.

**Cavity finesse measurement setup and method:**

The mirrors are mounted on six-axis translation stages for alignment. One of the mirrors is attached to a piezoelectric transducer (Thorlabs PC4RG15M) to modulate the cavity length. The tip and tilt of the mirrors are initially aligned parallel to each other, and the mode matching to the $TEM_{00}$ mode is optimized by adjusting the mirror positions.

To align the cavity, we use a retroreflector to reflect a collimated beam from the back mirror. Next, we introduce the front mirror and reflect light from its polished silicon backside. Finally, we confirm proper alignment by sending white light through the back of the cavity, ensuring it passes through both mirror apertures without clipping. More optimal alignment can be found by minimizing the birefringence of the cavity, or similarly, the splitting between degenerate modes like the $TEM_{01}$ and $TEM_{01}$.

When measuring finesse ($F$) and quality factor ($Q$) as a function of cavity length, we observe that changes in the mode waist affect mode matching. Additionally, reducing the finesse decreases transmission and weakens the signal. To compensate, we realign the mode-matching lens and increase the optical power to maintain a similar level of transmitted light.

As the cavity length increases, the swept ringdown spectroscopy method becomes less sensitive. The free spectral range (FSR) decreases with increasing length, causing the resonance frequency to vary more slowly at a given piezo sweep speed. To maintain a robust swept ringdown signal, the frequency sweep must exceed the ringdown time of the cavity; however, increasing the sweep speed is limited by the loaded bandwidth of the piezo when the cavity is attached.

For our measurements, we use an Eagleyard 780 nm mini-ECL laser (EYP-ECL-0780-00080-1500-BFW01-0005), frequency-locked to a rubidium reference resonance via modulation transfer spectroscopy[1]. As the piezo transducer modulates the cavity length, the resonance sweeps through the laser frequency, and the resulting ringdown can be observed in the time domain when the cavity shifts off resonance and the light decays. Fringes appear in the ringdown due to the high $F/Q$, where light circulates in the cavity on a timescale comparable to the length sweep.

$$I \propto e^{-\kappa t} |erfc(\frac{(i+1)}{2} \frac{\kappa/2 - i\alpha(t-t_0)}{i\sqrt{\alpha}})|^2$$

By fitting these fringes, we can extract the rate of frequency change across the resonance, $\alpha$. With this measurement in addition to a measurement of the amount of time required to travel a full FSR, $\Delta t_{FSR}$, we can extract the length of the cavity and the finesse.

$$\alpha = \frac{\nu_{FSR}}{\Delta t_{FSR}} = \frac{c}{2L \Delta t_{FSR}}$$

$$L = \frac{c}{2\alpha \Delta t_{FSR}} \quad F = 2\pi \frac{\alpha \Delta t_{FSR}}{\kappa}$$

Overall, this configuration enables accurate measurement of finesse and quality factors across a range of cavity lengths, even very short ones. To corroborate the results of the measurement, we calculate the length of the cavity by measuring the relative frequency spacing of the TEM modes. The resonance of a given TEM$_{mnq}$, where $q$ denotes the longitudinal mode and $m$ and $n$ denote the transverse mode indexes, is given by:

$$\nu_{mnq} = \frac{c}{2L}\left(q + \frac{1}{\pi}(m + n + 1)\arccos(\pm\sqrt{g_1 g_2})\right)$$

Where $g_i$ is the stability parameter given by, $g_i = 1 - \frac{L}{ROC_i}$. By measuring the spacing of these modes in time, we can get a direct translation between time and frequency.

$$\nu_{00q} - \nu_{00q+1} = \frac{c}{2L}$$

$$\nu_{01q} - \nu_{00q} = \frac{c}{2\pi L}\arccos(\pm\sqrt{g_1 g_2})$$

$$\frac{\nu_{01q} - \nu_{00q}}{\nu_{00q} - \nu_{00q+1}} = \frac{1}{\pi}\arccos(\pm\sqrt{g_1 g_2})$$

If we assume symmetric mirrors, this equation simplifies:

$$L = ROC\left(1 - \pi\frac{\nu_{01q} - \nu_{00q}}{\nu_{00q} - \nu_{00q+1}}\right)$$

We can then utilize the information from the profile measurements to extract a length from the measurements of the mode frequency spacing.

We also observe slight birefringence in the cavity, manifested by low-frequency beating in the transmission signal. This birefringence is typically on the order of the cavity linewidth, indicating a minor deviation from perfect symmetry, which could stem from imperfect alignment. To mitigate its effects, we align the input polarization to one of the cavity's two polarization modes.

Because the glued microcavity samples cannot be measured by sweeping their length, we perform swept ringdown spectroscopy by modulating the laser current. For the short glued microcavity, we use a Velocity™ TLB-6700 Widely Tunable Laser; rapid current modulation combined with piezo modulation allows us to sweep across the resonance quickly enough to observe the ringdown. We also leverage the laser's broad tuning range to measure a full FSR of the short glued microcavity, confirming its length and finesse. The accuracy of this FSR measurement depends on the internal calibration of the Velocity 6700.

For the longer glued microcavity, its narrow resonance lets us use the same Eagleyard 780 nm mini-ECL employed for unglued samples. Because the laser current can be swept sufficiently fast relative to the cavity linewidth, swept ringdown spectroscopy remains feasible with this setup.

**FEM simulation of the buckling mode**

COMSOL is used to simulate the shape of the resulting mirror coatings when subject to compressive stress from the deposition process. This allows us to accurately predict their form apart from the characterization of fabricated devices. Since all mirrors have spherical shapes, we apply the axial symmetry condition to speed up the simulation, and model a disk shape that is hard-clamped at the perimeter boundary. The disk consists of 21 alternating layers of $SiO_2$ and $Ta_2O_5$ along the axial direction, with $SiO_2$ at 134 nm thickness, $Ta_2O_5$ at 94 nm thickness for each layer, and a disk radius of varying sizes.

We begin with a uniaxial compressive strain $\epsilon = 5 \times 10^{-4}$ across the material stack and perform a Linear Buckling study to determine the critical load factor, which provides the threshold compressive strain required for buckling. Using the buckling profile from this study as an initial condition, we then conduct a stationary study that sweeps through a range of strains where buckling occurs. This stationary study generates multiple buckling profiles corresponding to different strain levels.

To anchor the strain in our device, we match the maximum displacement in the simulation with measurement results from an optical profiler. With the strain $\epsilon = 3 \times 10^{-3}$ obtained this way, we find that the simulated buckling profiles closely match the measured data from fabricated devices.

**Other micromirror geometries**

As shown in Fig. 4, we create micromirrors that can be glued together to form microcavities. With the same inner diameters, which determine the depth and ROC, the outer diameter can also take different values without affecting the mirror characteristics, to satisfy different space constraints. Some examples are shown in Fig. S2 (b).

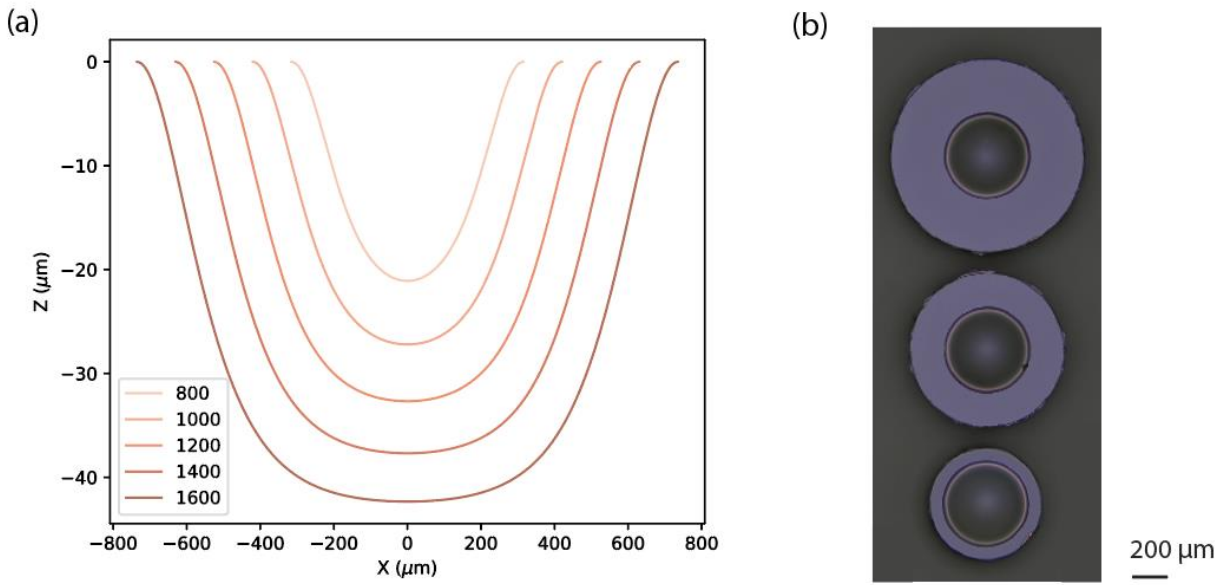

Fig. S2. Various mirror geometries: (a) The simulated profiles of the buckled mirrors overlaid for different diameters of the mirror. They are the data points used to extrapolate depth and ROC in Fig. 2(D). The labels correspond to the mirror diameter in μm. (b) A microscope image of micromirrors of the same inner radius (400 μm) and different outer radii.